\documentstyle[floats,aps,epsf]{revtex}

\newif\ifsmallfigures
\smallfigurestrue
\begin{document}
\twocolumn[\hsize\textwidth\columnwidth\hsize\csname @twocolumnfalse\endcsname

\title{Isolated resonances in conductance fluctuations and hierarchical
  states} 

\author{Arnd B\"acker$^{1,2}$, Achim Manze$^3$, Bodo Huckestein$^3$,
  and Roland Ketzmerick$^4$}

\address{$^1$  School of Mathematics, University of Bristol,
  University Walk, Bristol BS8 1TW, UK and
   BRIMS, Hewlett-Packard Laboratories, Filton Road, Bristol
  BS12 6QZ, UK\\ 
  $^{2}$ Abteilung Theoretische Physik, 
  Universit\"at Ulm, Albert-Einstein-Allee 11, D-89081 Ulm, Germany\\ 
  $^{3}$ Institut f\"ur Theoretische Physik III, Ruhr-Universit\"at
  Bochum, D-44780 Bochum, Germany\\ 
  $^{4}$ Max-Planck-Institut f\"ur Str\"omungsforschung and Institut
  f\"ur Nichtlineare Dynamik der Universit\"at G\"ottingen,\\ 
  Bunsenstra{\ss}e 10, D-37073 G\"ottingen, Germany}

\date{January 29, 2002}

\maketitle

\begin{abstract}
  We study the isolated resonances occurring
  in conductance fluctuations of
  quantum systems with a classically mixed phase space.  We
  demonstrate that the isolated resonances and their
  scattering states can be associated to eigenstates of the closed
  system. They can all be categorized as
  hierarchical or regular, depending on where the 
  corresponding eigenstates live in the
  classical phase space.
\end{abstract}
\vskip2pc

]

\section{Introduction}
\label{sec:introduction}

The classical dynamics of a scattering system is
reflected in the transport properties of its quantum mechanical
analog.  A prominent example in quantum chaos are the
universal conductance fluctuations exhibited by a scattering system
with classically completely chaotic dynamics~\cite{Jal2000}.
Generic systems, however, are neither
completely chaotic nor integrable, but show chaotic as well as
regular motion~\cite{LicLie92}.
The chaotic dynamics is strongly influenced by the presence of islands 
of regular motion, in particular, one finds a trapping of chaotic 
trajectories close to regular regions with trapping times distributed 
according to power laws~\cite{ChiShe81}.
The semiclassical analysis revealed that conductance fluctuations of 
generic scattering systems have corresponding power-law 
correlations~\cite{LaiBluOttGre92,Ket96}
and most interestingly that the graph conductance vs. control 
parameter is a fractal \cite{Ket96}. 
Fractal conductance fluctuations have indeed been observed
experimentally in semiconductor 
nanostructures~\cite{SacKetGouFenKelDelWas98,Tay2001}
as well as numerically~\cite{CasGuaMas2000}.

Surprisingly, for the cosine billiard\cite{Stifter,cosinus}, a system
with a mixed phase space and power-law distributed classical trapping
times, a recent numerical study did not show fractal conductance
fluctuations~\cite{HucKetLew99}.  Instead, sharp isolated resonances
were found with a width distribution covering several orders of
magnitude.  Only about one third of them can be related to quantum
tunneling into the islands of regular motion~\cite{Seb93}, while the
rest remained unexplained.  It was later shown that conductance
fluctuations for mixed systems should in general show fractal
fluctuations on large scales and isolated resonances on smaller
scales~\cite{HufWeiKet2001}.  The isolated resonances in the
scattering system were conjectured to be related to a subset of
eigenstates of the closed system, namely hierarchical
states~\cite{KetHufSteWei2000} living in the chaotic component close
to the regular regions and regular states living within the islands of
regular motion~\cite{Seb93}.  This type of behavior was obtained for a
quantum graph which modeled relevant features of a mixed phase
space~\cite{HufWeiKet2001}.

The purpose of the present paper is to establish the origin of all
isolated resonances for a system with a mixed phase space.
To this end, we study the cosine billiard for suitable parameters
in a three-fold way:
(i) as a quantum scattering system, (ii) as a closed quantum system,
and (iii) its classical phase-space structures.
We find that the resonances have scattering states and corresponding
eigenstates of the closed system that live
in the hierarchical and regular part of phase space.
The number of resonances of each type is directly related to
the corresponding volumes in the classical phase space.
Each resonance width is quite well described by the strength of the
corresponding eigenfunction at the billiard boundary. 
Exceptions are shown to arise from the presence of avoided crossings 
in the closed system.
It is demonstrated that the simultaneous appearance of
fractal conductance fluctuations and isolated resonances,
as observed in a quantum graph model~\cite{HufWeiKet2001},
would for our system with a mixed phase space
require much higher energies. These are
currently computationally inaccessible.

In the following section, we define the model 
we use to study the relation between the scattering resonances and
the eigenstates of the corresponding closed system.
Our main results on the classification of resonant scattering states and 
corresponding eigenstates of the closed system into hierarchical and regular
are presented in Secs.~\ref{sec:scatter} and \ref{sec:closed}.
The role of partial transport barriers is analyzed in
Sec.~\ref{sec:barriers}. In Sec.~\ref{sec:avoidedcrossings} we discuss
the effect of avoided crossings on the assignment of resonances of the open billiard to
eigenstates of the closed system and Sec.~\ref{sec:conclusion} gives a
summary of the results. Finally, the Appendix contains some details of the
numerical methods employed in the present work.

\section{The Model}
\label{sec:model}

\begin{figure}[tbp]
  \begin{center}
    \epsfxsize=8.3cm
    \leavevmode
    \epsffile{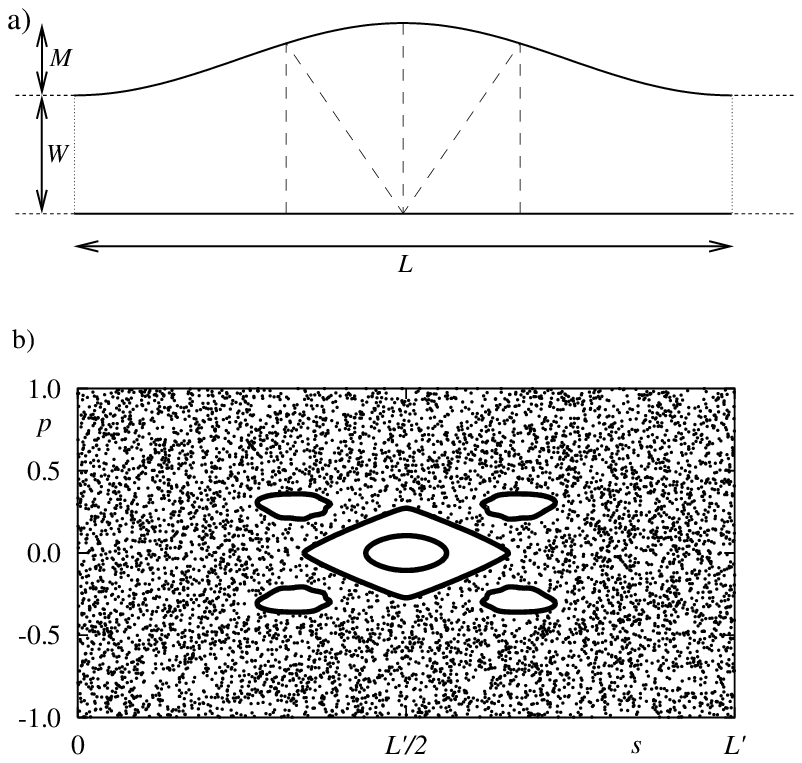}

    \caption{(a) The cosine billiard with semi-infinite leads (short dashed
      lines) and hard walls for closing the system (dotted lines) for
          $W/L=0.18$ and $M/L=0.11$. Also shown are the two most prominent 
          stable periodic orbits for these parameters (long dashed lines).
          (b) Poincar\'e section of  some regular and one chaotic 
          orbit
           for the above parameters in
          Poincar\'e-Birkhoff coordinates $p$ vs arclength 
         $s$ along the upper
          boundary of the billiard. 
          A major island at $(s,p)=(L'/2,0)$
          around the elliptic I-shaped orbit and
          4 smaller islands surrounding the M-shaped orbit can be seen.}
    \label{fig:cosine}
  \end{center}
\end{figure}

We study the cosine billiard \cite{Stifter,cosinus}, either closed or with
semi-infinite leads attached. The boundaries of the billiard are hard
walls (i.e. Dirichlet boundary conditions) at $y=0$ and
\begin{equation}
  \label{eq:1}
  y(x) = W + \frac{M}{2}\left( 1 - \cos\left(\frac{2\pi
        x}{L}\right)\right), 
\end{equation}
for $0\leq x \leq L$ (see Fig.~\ref{fig:cosine}a). In the open billiard
two semi-infinite leads of width $W$ are attached at the openings at
$x=0$ and $x=L$, while in the closed billiard the openings are closed
by hard walls. The classical phase-space structure
can be changed by varying the ratios $W/L$ and $M/L$. 
For $M/L=0$ the dynamics is integrable 
and for example for $M/L=1/2$ and $W/L=1$  the dynamics appears to be ergodic
(at least the islands of regular motion, if any, are very small) \cite{Stifter}.

In the present work, we use the same parameters as in Ref.~\cite{HucKetLew99}, 
namely $W/L=0.18$ and $M/L=0.11$, for which the I- and M-shaped orbits
depicted in Fig.~\ref{fig:cosine}a are stable.
The corresponding Poincar\'e section is shown in 
Fig.~\ref{fig:cosine}b.
We use Poincar\'e-Birkhoff coordinates $(s,p)$, where $s$ is
the arclength along the upper boundary of the billiard 
with length
$L'\approx1.029\,L$ and $p$ is the projection of the unit momentum vector after
the reflection on the tangent in the point $s$.

Quantum mechanically, for a given wave number $k_F$ the number $N$ of
transmitting modes in a lead of width $W$ is the largest integer with
$N \leq k_F W/\pi$.  We measure energies in units of the energy
$E_0=\hbar^2 \pi^2 / (2 m W^2)$ of the lowest mode in such a lead,
i.e., $E=\hbar^2 k_F^2/(2mE_0)=(k_F W/\pi)^2 \geq N^2$.  The larger
the number $N$ of modes is, the more details of the classical phase
space can be resolved by quantum mechanics.  At the same time the
computational effort increases with $N^4$ and we compromise, as in
Ref.~\cite{HucKetLew99}, on the case of $N=45$ transmitting modes in
the energy range $E\in [2026,2100]$.

\section{Resonances and scattering states}
\label{sec:scatter}

Resonances in the scattering system, which have been observed as isolated features
in conductance fluctuations~\cite{HucKetLew99}, were identified by isolated peaks
in the Wigner-Smith time delay $\tau$ of the system. The time delay is
given by
\begin{equation}
  \label{eq:2}
  \tau=\frac{-i\hbar}{2N} \mbox{Tr}\,(S^\dagger dS/dE),
\end{equation}
where $2N$ is the dimension of the $S$-matrix. The calculation of $S$
and $\tau$ was already outlined in Ref.~\cite{HucKetLew99} and is presented
in greater detail in Appendix~\ref{sec:hybr-repr-recurs}.

\begin{figure*}[tbp]
  \begin{center}
    \epsfxsize=17cm 
    \leavevmode
    \epsffile{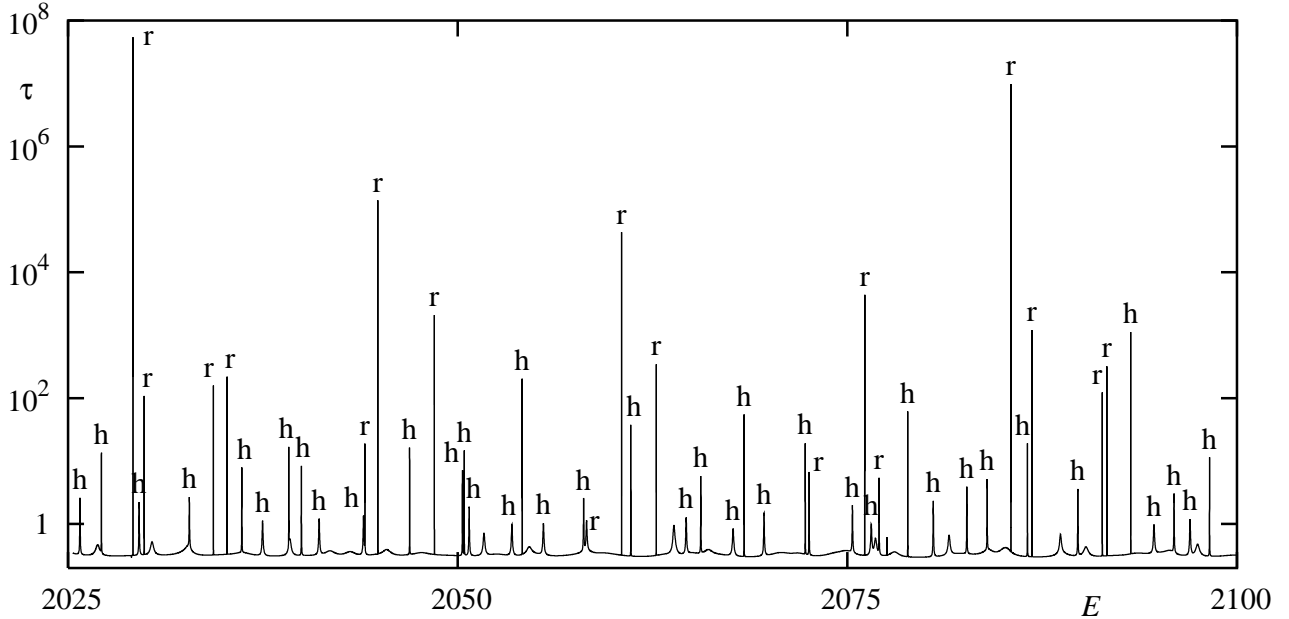}
    \vspace*{2ex}

    \caption{Wigner-Smith delay time $\tau$ vs energy $E$. 
        For each resonance a corresponding eigenstate of the closed
        system was found and the labels indicate whether it lives in the
        regular ($r$) or hierarchical ($h$) region of phase space.}
    \label{fig:tau}
  \end{center}
\end{figure*}

In Fig.~\ref{fig:tau} we show the Wigner-Smith time delay $\tau$ [in
units of $\hbar/E_0=2mW^2/(\hbar\pi^2)$] for $E\in[2026,2100]$.
The isolated resonances found in Ref.~\cite{HucKetLew99} are
clearly seen. 
Each isolated resonance $E_{{\rm res},i}$ has a Breit-Wigner shape
\begin{equation}
  \label{eq:bw}
  \tau_i(E) = \tau_i \,\frac{\Gamma_i^2/4}{(E-E_{{\rm res},i})^2+\Gamma_i^2/4},
\end{equation}
with $\tau_i\Gamma_i=2/N$. Note that the heights $\tau_i$ and the
corresponding widths $\Gamma_i$ of the
individual resonances cover several orders of magnitude.

In order to elucidate the nature of the resonances, we calculated the
scattering states inside the open billiard. 
For a given configuration of waves incoming in both leads, the                
knowledge of the $S$-matrix allows the determination of the outgoing          
waves and hence the wavefunction amplitudes at the openings of the            
billiard.                                                                     
Since the $S$-matrix is defined between
asymptotic, propagating modes, this procedure neglects the
contribution of evanescent modes in the leads in the vicinity of the
billiard. The wavefunction amplitudes at the openings can then be used as
boundary conditions for the solution of the Schr\"odinger equation
inside the billiard. For the examples of scattering states presented
below, we  occupied the 10 topmost modes incoming from the left 
with equal amplitudes. 
Similar pictures were obtained for other boundary
conditions.

For the comparison of the scattering states
with the classical phase-space structures
we have calculated Husimi projections $H^{sc}(s,p)$.
Similar to the case of closed billiards, see sec.~\ref{sec:closed},
we define these by the projection of the
scattering state onto a coherent state on the upper boundary of
the billiard,
\begin{eqnarray}
  \label{eq:10}
  H^{sc}(s,p) &=& |\langle 
\partial_{{\bf n}}\psi| \phi^{\text{coherent}}_\gamma
  (s,p)\rangle|^2 \\
  &=&
  \left|\int^{L'}_0 ds' \partial_{{\bf n}}\psi^*(s') e^{i k p (s'-s) -
  \frac{1}{2} k (s'-s)^2} \right|^2,
\end{eqnarray}
with $k=\sqrt{E}\pi/W$. Here $\partial_{{\bf n}}\psi(s)=
{\bf n}(s) \cdot {\bf\nabla}\psi({\bf q}(s))$ is the
normal derivative of the scattering state on the upper boundary. 
${\bf n}(s)$ is the normal vector and ${\bf q}(s)$ is the position of the 
boundary as a function of arc length $s$.
Note that these Husimi projections are not normalized and are
influenced by the openings over a range of a few Fermi wavelengths.
Also they do not include the full billiard boundary
and therefore no periodization of the coherent state
has been used.

\begin{figure*}[tbp]
  \begin{center}
    \epsfxsize=17cm
    \leavevmode
    \ifsmallfigures\epsffile{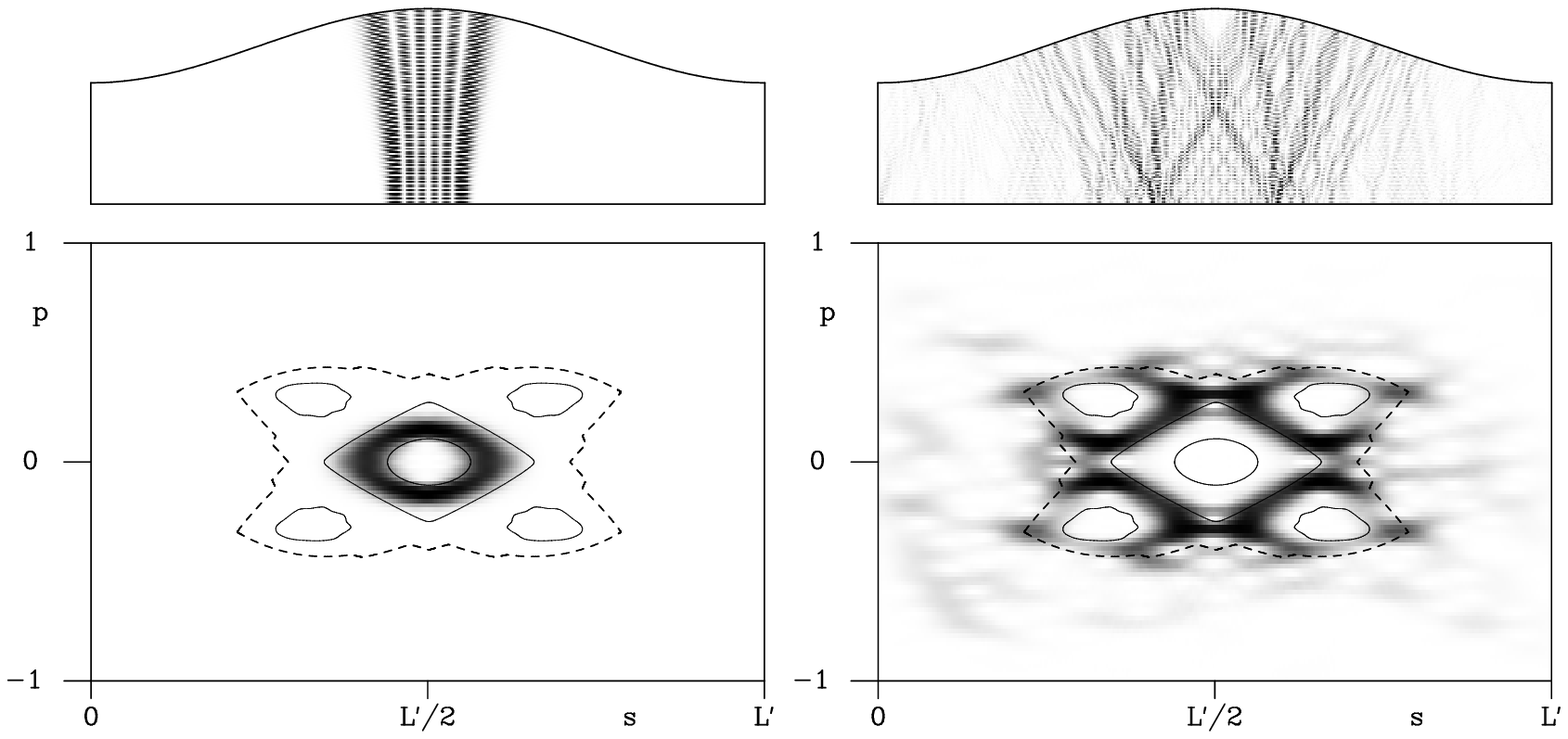}\else\epsffile{figure3.ps}\fi
    
    \caption{Resonant scattering states (top row) and Husimi
      projections (bottom row)
          onto the classical Poincar\'e section with
      KAM-tori (solid lines) and a partial transport
      barrier (dashed line).
          Two example at resonance energies
      2029.172 (left) and 
          2041.109 (right) are shown. They live in the 
      regular and the hierarchical
          region of phase space, respectively.
          For the representation of the scattering states a superposition with equal weight of the 10 topmost modes incoming from the left is shown.         
}
    \label{fig:scatter_examples}
  \end{center}
\end{figure*}

As a first example, we present in Fig.~\ref{fig:scatter_examples} on
the left the scattering state at an energy of approximately 2029.172,
the center energy of the sharpest observed resonance. Obviously, the
scattering state is associated with the I-shaped periodic orbit. The
wavefunction amplitude is concentrated near the orbit and the Husimi
projection lives predominantly inside the central stable island of the
classical phase space. For comparison, we present in
Fig.~\ref{fig:scatter_examples} on the right the scattering state at
energy 2041.109. The width of the resonance at this energy is about
$6\cdot10^7$ times larger than the width of the sharpest resonance.
Evidently, this resonance is not related to the stable islands in
phase space. In contrast, by comparing with the superimposed KAM-tori
of the Poincar\'e section and a partial transport barrier
surrounding the island hierarchy (see
Section~\ref{sec:barriers}), we see that the Husimi projection
lives in the hierarchical region 
between the islands and a partial transport barrier.

As scattering states allow a great variability in the
boundary conditions, e.g., the incoming modes,
we do not use them for a detailed analysis of the isolated resonances.
Instead we consider the corresponding eigenstates of the
closed system in the next section.

\section{Resonances and corresponding eigenstates}
\label{sec:closed}

In this section we want to demonstrate that the isolated resonances of
the conductance fluctuations and their scattering states have
corresponding eigenstates of the closed billiard. In particular,
we will show that {\em all} these eigenstates live in the hierarchical and
regular part of phase space, as was conjectured in Ref.~\cite{HufWeiKet2001}.
This allows a labeling of all isolated resonances
appearing in Fig.~\ref{fig:tau}.

For the closed system the eigenvalues and eigenfunctions are computed
using the boundary element method, see, e.g.~\cite{Bae2002:p} and
references therein.
Since the cosine billiard is symmetric
with respect to the axis $x=L/2$, the eigenstates have definite parity
$P=+,-$. The actual calculations are performed for the desymmetrized
billiard with either Dirichlet or Neumann boundary conditions on
the symmetry axis yielding the antisymmetric ($P=-$) and symmetric
($P=+$) states, respectively. We label the $n$-th eigenstate of
parity $P$ by $n^P$.
The mean level spacing $\Delta$ is determined by the area 
$A=L(W+M/2)$ of the billiard using Weyl's formula
$\Delta/E_0 = (4\pi\hbar^2/2mA)/E_0 = 0.176$.

\begin{figure*}[tbp]
  \begin{center}

    \epsfxsize=17cm
    \leavevmode
    \ifsmallfigures\epsffile{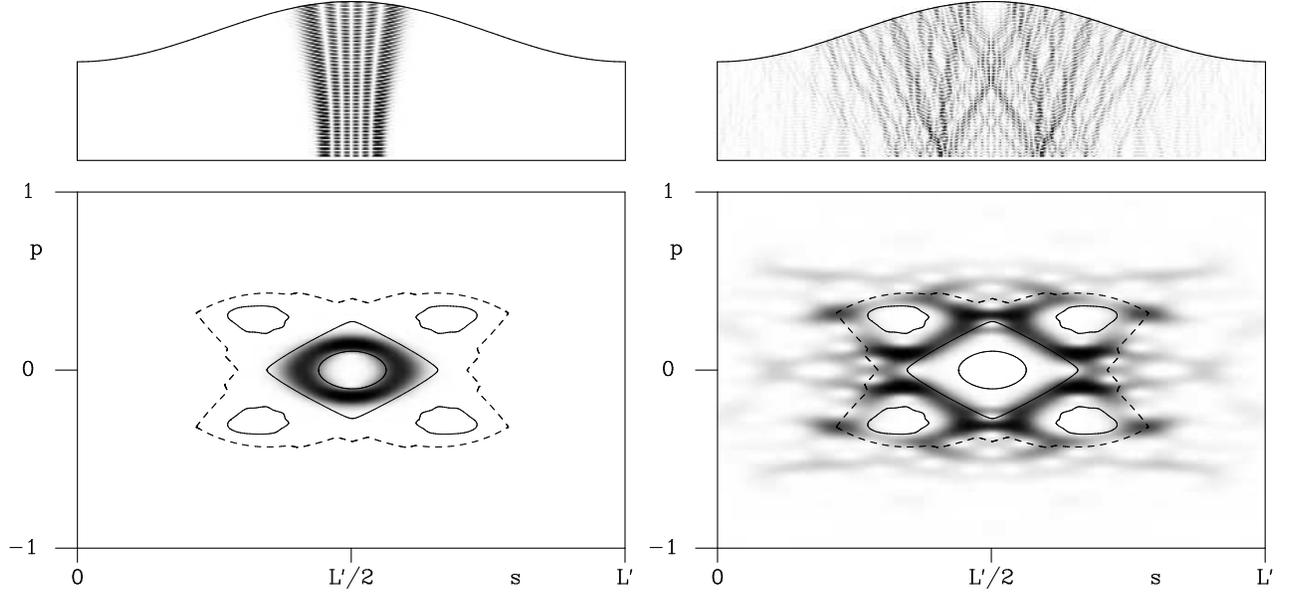}\else\epsffile{figure4.ps}\fi
    \vspace*{2ex}

    \caption{Eigenfunctions (top row) and Husimi representations
      (bottom row) of a regular state ($5686^-$, left) and a
      hierarchical state ($5720^-$, right)
          corresponding to the scattering states of Fig.~\ref{fig:scatter_examples}.
      }
    \label{fig:husis}
  \end{center}
\end{figure*}

We present in Fig.~\ref{fig:husis} the two eigenstates
corresponding to the scattering states shown in
Fig.~\ref{fig:scatter_examples}. For each state, we show the
eigenfunction density $|\psi_n({\bf q})|^2$ and the corresponding
Husimi representation $H_n(s,p)$ (see e.g.,~\cite{TuaVor95,SimVerSar97}).  The state $5686^-$, displayed on
the left of Fig.~\ref{fig:husis} differs in energy by about $0.01 \Delta$ from the sharpest observed resonance with
energy 2029.172. Note that this energy difference is of the order of
the accuracy to which our resonance energies and eigenvalues are
calculated. On the right hand side of Fig.~\ref{fig:husis}, a
hierarchical state is displayed. Its energy differs from the resonance
at energy 2041.109 by about $0.1\Delta$. This shift of the resonance energy from the eigenenergy of the closed system is due to the opening of the system by attaching the leads. As for the
scattering states, we have superimposed some KAM-tori onto the Husimi
representations of Fig.~\ref{fig:husis}.
In addition, a partial transport barrier
surrounding the island hierarchy is shown (see next section).

\begin{figure}[!b]
  \begin{center}
    \epsfxsize=8.6cm
    \leavevmode
    \epsffile{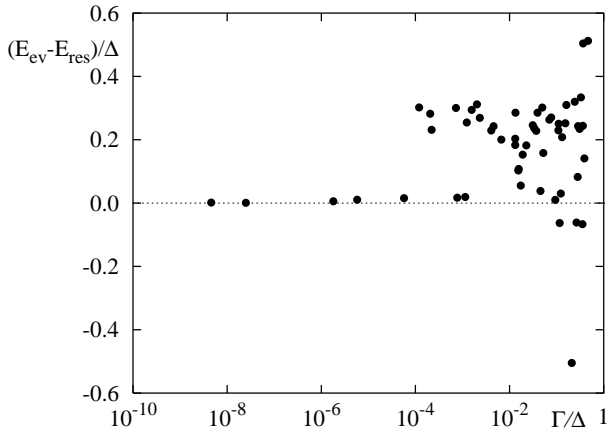}
    \vspace*{2ex}

    \caption{Difference of eigenstate energy $E_{\rm ev}$
        and resonance energy $E_{\rm res}$ in units of the mean
        level spacing $\Delta$ vs $\Gamma/\Delta$.
        The deviations increase with $\Gamma$.}
    \label{fig:DeltaE-vs-Gamma}
  \end{center}
\end{figure}

\begin{figure}[!b]
  \begin{center}
    \epsfxsize=8.6cm
    \leavevmode
    \epsffile{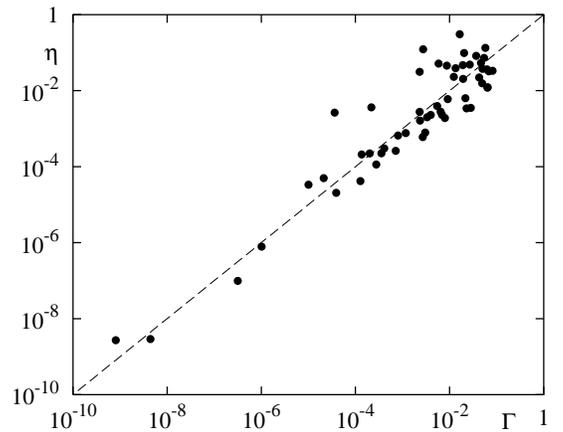}
    \vspace*{2ex}

    \caption{The strength $\eta$ of an eigenstate at the
        left boundary vs the resonance width $\Gamma$
        of the corresponding resonance.
        An approximate proportionality can be seen.}
    \label{fig:eta-vs-Gamma}
  \end{center}
\end{figure}

Now we want to associate all resonances of the scattering system with width
$\Gamma$ at energy $E_{\rm res}$ with an eigenstate of the closed
billiard with energy $E_{\rm ev}$. 
We use a Husimi representation $H_n(s,p)$ on the 
Poincar\'e section to determine the region on which an eigenstate localizes.
We introduce the quantity
\begin{equation}
  \label{eq:18}
  \eta_n = 
           \int_{-W}^{0} ds \int_{-1}^{1} dp \, H_n(s,p) ,
\end{equation}
which integrates the Husimi distribution over the left boundary 
of the billiard (not shown in Fig.~\ref{fig:husis})
with the normalization of the Husimi distribution chosen such that
$\int_{-W}^{L'/2} ds \int_{-1}^{1} dp \, H_n(s,p) =1$.
This quantity gives an estimate
on how strongly a state of the closed system will couple to the leads in
the scattering system and should be roughly proportional to $\Gamma$. 
This allowed us to find, for each of the 54 resonances with
$\Gamma \leq \Delta/2$, a state with 
$E_{\rm ev} \approx E_{\rm res}$ and with $\eta \approx \Gamma$.
Figures~\ref{fig:DeltaE-vs-Gamma}~and~\ref{fig:eta-vs-Gamma} show
the difference $E_{\rm ev}-E_{\rm res}$ in units of the mean
level spacing $\Delta$ and the approximate
proportionality of $\eta$ and $\Gamma$, respectively.
Clearly the larger differences appear for bigger $\Gamma$, but
still a clear identification is possible 
(see section~\ref{sec:avoidedcrossings}).
This assignment also works the other way around, as
of the 46 eigenstates with the smallest values of $\eta$, we can 
identify 40 with isolated resonances, missing only the 6 regular 
eigenstates quantized most deeply in the central island of phase 
space, as discussed below.

For the 54 resonances with width $\Gamma$ less than half a mean level
spacing, we analyze the structure of the corresponding eigenstates. We find
that 17 states can be categorized as regular states, as their Husimi
representations live inside the five major stable islands in phase
space. Of these states, 7 are associated with the I-shaped orbit and
10 with the M-shaped orbit. While we observe all states in the energy
interval associated with the M-shaped orbit, 6 further eigenstates
living near the center of the central stability island are not
resolved as resonances. As these are the innermost states in the
island, we expect them to couple weaker to the leads and their
resonance widths to be much smaller than the sharpest observed
resonance. Apparently, these resonances are so narrow that we were not
able to find them, given our current numerical accuracy, even knowing
their approximate energy from the eigenvalues.

The remaining 37 resonances are not related to regular states, but
the Husimi representations of their corresponding eigenstates
have large intensity in the region between the regular islands and the
partial transport barrier and a much weaker intensity in the rest of
the chaotic region.
It should be noted that in the studied energy range accessible to
our methods the wave length is of the order of the distance between
regular islands and the partial transport barrier. 
Therefore the eigenstates either look like regular states living
outside the island~\cite{BohTomUll93} or similar to  scarred
states on a hyperbolic orbit close to the island~\cite{RadPra8891}. 
For higher energies they would
show the true properties of hierarchical
states, i.e., being similar to a chaotic state, but restricted to 
the hierarchical region~\cite{KetHufSteWei2000}.
We therefore classify these states as hierarchical states.

In Fig.~\ref{fig:tau} we have labeled the resonances by $r$ and $h$
according to our classification of the corresponding eigenstates
as regular and hierarchical, respectively. 
This demonstrates that the origin of {\em all\/} isolated resonances
are hierarchical or regular eigenstates of the closed system.

\section{Partial transport barriers}
\label{sec:barriers}

Classical transport in the chaotic part of phase space is
dominated by partial 
barriers~\cite{KayMeiPer84,HanCarMei85,MeiOtt86,GeiZacRad87,Mei92}.
They are formed by cantori as well as by stable and unstable
manifolds.
Such a partial transport barrier coincides with its iterate, with
the exception of so-called turnstiles where phase-space volume is
exchanged between both sides of the partial barrier.
We have constructed partial barriers using the 
methods described in Ref.~\cite{KayMeiPer84}.
The fluxes $\Phi$ are determined from the length $l$ of the maximizing
and minimax orbits, according to
\begin{eqnarray}
  \Phi &=& \hbar k_F |l_{\rm maximizing} - l_{\rm minimax}| \\
       &=& \hbar \pi \sqrt{E} |l_{\rm maximizing} - l_{\rm minimax}|/W .
\end{eqnarray}

Quantum mechanically, partial transport barriers with fluxes up to the order of $\hbar$
divide the chaotic part of phase space into 
distinct regions with chaotic and hierarchical eigenfunctions living mainly on one or the
other side~\cite{KetHufSteWei2000}.
We found that the partial barrier with smallest flux that surrounds
the main island and the 4 neighboring islands can be constructed
from the stable and unstable manifolds of the period 4 hyperbolic
fixed points.
Each of its two turnstiles has for our largest energy $E=2100$ a
flux $1.06 \hbar$.
Further outside are many other partial barriers with only slightly
bigger fluxes. 
As an example, we show in Figs.~\ref{fig:scatter_examples} and
~\ref{fig:husis} the partial barrier
constructed from an unstable periodic orbit with winding number
$5/23$, which is an approximant of the most noble irrational between
winding numbers $1/4$ and $1/5$.
It has a flux $1.65 \hbar$.

A check on the validity of our identification of regular and
hierarchical states is provided by a comparison of their numbers 
to the corresponding relative volumes in phase space. To this end, we calculate
the volume of the tori associated with stable periodic orbits, $V_r$,
and the chaotic phase-space volume inside the partial transport 
barrier, $V_h$. We find $V_r$ and $V_h$ to
cover 5.9\% and 8.5 \% 
of the energy shell, respectively. From the
total number of eigenstates in the energy interval, $N=426$, we get for
23 (17+6) regular and 37 hierarchical states
relative fractions of 5.4\% and 8.7\%, respectively, in good 
agreement with the volumes of the associated regions in phase space.

The absence of fractal conductance fluctuations in this system has
now a clear explanation:
According to Ref.~\cite{HufWeiKet2001} for fractal fluctuations to occur
a hierarchy of partial transport barriers with fluxes
larger than $\hbar$ must exist.
For the studied energies
we find that even the outermost partial barriers surrounding
the hierarchical phase-space structure have fluxes of the order
of $\hbar$.
This causes a quantum dynamical decoupling of 
the chaotic part connected to the leads from the
entire hierarchical part. As the hierarchical region 
is the semiclassical origin of fractal fluctuations,
they are not observed.
For much higher energies only, the hierarchy of partial transport barriers
would have an outer region with fluxes larger than $\hbar$, leading
to fractal conductance fluctuations.
The inner region of this hierarchy with fluxes smaller
than $\hbar$ has now a smaller phase-space volume.
Still, together with the regular regions it will cause isolated resonances
on smaller scales than the fractal fluctuations.
Unfortunately, this energy regime is currently computationally inaccessible
for the studied system.

Power laws in the distribution of resonance widths and in
the variance of conductance increments had been observed in
Ref.~\cite{HucKetLew99}, apparently reflecting the classical
dwell time exponent. They were related to the resonances below the
mean level spacing. For these resonances we now have demonstrated that they
arise due to hierarchical and regular states.
This allows to apply the arguments of Ref.~\cite{HufWeiKet2001} about the
resonance width distribution of hierarchical states. They lead to the
conclusion that these apparent power laws come from broad transition regions to
asymptotic distributions that are unrelated to the classical dwell time
exponent.

\section{Avoided crossings}
\label{sec:avoidedcrossings}

\begin{figure}[tbp]
  \begin{center}

    \epsfxsize=8.6cm
    \leavevmode
    \ifsmallfigures\epsffile{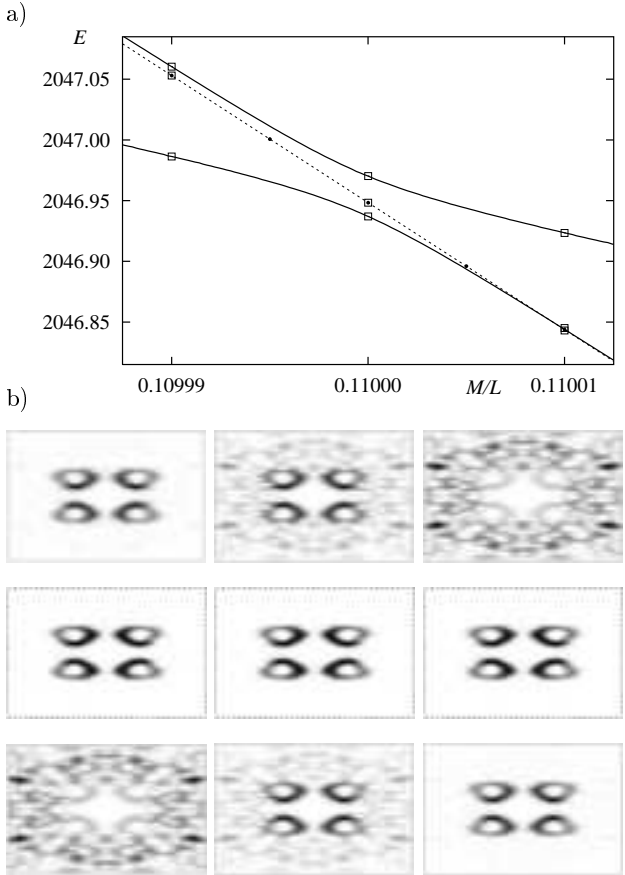}\else\epsffile{figure7.ps}\fi
    \vspace*{2ex}

     \caption{(a) Energies of states $5736^-$ and $5737^-$  (solid
       lines) showing an avoided crossing under variation of $M/L$.
       The energy of the only isolated resonance in this energy range
       (dots connected by a dashed line) follows the regular state of
       the closed system.  The slight offset in the resonance energy
       is within the systematic numerical error of the numerical
       method.  (b) The Husimi representations for state $5737^-$ (top
       row), the
       scattering state (middle row) and for eigenstate $5736^-$
       (bottom row) are shown for $M/L=0.10999,\,0.11,\,0.11001$ (left
       to right). For the
       eigenstates one clearly sees the typical exchange of the
       structure while passing the avoided crossing whereas the
       scattering state is not affected.}
    \label{fig:paramvar}
  \end{center}
\end{figure}

While for most resonances and corresponding eigenstates the parameters
$\eta$ and $\Gamma$ are of the same order of magnitude, for some
states $\eta$ exceeds $\Gamma$ by up to 2 orders of magnitude.  This
phenomenon can be understood as an effect of avoided level crossings
in the closed system: In Fig.~\ref{fig:paramvar}a) we show as an
example the dependence of the energy of states $5736^-$ and $5737^-$
as a function of the parameter $M/L$ for the narrow range $0.10999\leq
M/L \leq 0.11001$, displaying the typical features of an avoided
crossing. A comparison of the associated Husimi representations shows
that the states $5736^-$ and $5737^-$ do indeed exchange their
character from chaotic to regular and from regular to chaotic,
respectively, showing a superposition at $M/L=0.11$. Upon opening the
system, the chaotic state couples much more strongly to the leads as
compared to the regular one.  Consequently, in the complex energy
plane of the scattering system there is no longer an avoided crossing.
The regular state leads to an isolated resonance with an almost linear
energy dependence on $M/L$ and the phase-space signature of the
regular state (middle row in Fig.~\ref{fig:paramvar}b). It closely
follows the expected energy dependence of the regular state in the
closed system if it had not made an avoided crossing with the chaotic
state (Fig.~\ref{fig:paramvar}a).

Another example of an avoided crossing is given by state $5801^-$
with an eigenvalue about $0.5\Delta$ less than the resonance position
(see lower right corner of Fig.~\ref{fig:DeltaE-vs-Gamma})
and the state $5802^-$, with an eigenenergy about
$1.4\Delta$ above the resonance energy. Both states show similar
Husimi representations and have $\eta$ values exceeding $\Gamma$
by about a factor of ten.

In all the cases when $\eta$ drastically exceeds $\Gamma$ a
closer look at these eigenstates reveals that they are superpositions
of regular or hierarchical states with chaotic states.
They are due to avoided crossings and the chaotic
part leads to a comparatively large value of $\eta$. In
contrast, in the open system no avoided crossing
occurs in the complex energy plane and the resonance width $\Gamma$ 
is unaffected.

\section{Conclusion and Outlook}
\label{sec:conclusion}

We demonstrate a clear correspondence of the isolated resonances
observed in the transport properties of the open cosine billiard to
hierarchical and regular eigenstates of the closed billiard.
We can identify all resonances
with widths less than half of the mean level spacing.
The classification of resonances into a hierarchical or regular
origin yields numbers in agreement with the relative phase-space volumes.
On a quantitative level, we find a roughly linear relation
between the widths of the isolated resonances and the weights
of the associated eigenstates at the part of the
boundary where the leads are attached. States with unusually large
weights can be attributed to 
avoided crossings with chaotic eigenstates. 

We find that the island hierarchy is
separated from the chaotic part of phase space by partial transport
barriers with fluxes of the order of $\hbar$.
This supports the notion that the absence of
fractal conductance fluctuations in the currently accessible energy range is due to the quantum
dynamical decoupling of the hierarchical part of phase space from the
chaotic part connected to the external leads.

The simultaneous appearance of isolated resonances and fractal fluctuations,
beyond the quantum graph model studied in Ref.~\cite{HufWeiKet2001},
remains to be demonstrated numerically or experimentally
for a system with a mixed phase space. Numerically, the challenge is the observation of fractal fluctuations of the
conductance, which go beyond one order of magnitude \cite{TakPlo00}. This requires
calculations with a drastically increased number of
modes, the use of improved techniques like the modular recursive Green's
function method \cite{RotTanWirTroBur00}, and the search for suitable billiard systems where the
turnstile fluxes across partial barriers are particularly large. Isolated
resonances will easily appear as soon as the parameter is varied on a
sufficiently small scale. We note, that fluctuations of the quantum staying probability, which can  be fractal~\cite{CasGuaMas2000},  cannot show isolated resonances.  Similarly, we expect no appearance of isolated resonances
within the fractal fluctuations observed in recent studies \cite{LouVer00etaliud}, as they are
unrelated to a classical mixed phase space.

On the experimental side, fractal conductance fluctuations
have been observed \cite{SacKetGouFenKelDelWas98,Tay2001} and also isolated resonances coming from regular regions have been recently reported \cite{dMoTBP}. The simultaneous appearance of both types
including isolated resonances from hierarchical regions requires to go far
enough into the semiclassical regime, i.e. to quantum dots with dimensions
bigger than 1 $\mu$m, as in Ref.~\cite{SacKetGouFenKelDelWas98}. At the same time the phase coherence time
must be large enough to resolve isolated resonances of a given width and, of
course, the parameter, typically a magnetic field, must be varied on a
sufficiently fine scale. Given the experimental limitations it would be helpful
if an optimal form for such a quantum dot could be provided by theoretical
considerations. This seems to be quite difficult at present, since the
difference in the lithographic shape and the actual potential experienced by
electrons has dramatic consequences on the electron dynamics and thus on the
scales over which fractal fluctuations and isolated resonances appear.

\vspace{1cm}

\noindent{\bf Acknowledgments}

\vspace{0.25cm}

\noindent
{R.K. acknowledges helpful discussions with
L.~Hufnagel and M.~Weiss. A.B.\ acknowledges support by the 
Deutsche Forschungs\-ge\-mein\-schaft under contract No. DFG-Ba 1973/1-1.

\appendix

\section{Hybrid representation and recursive Green function method}
\label{sec:hybr-repr-recurs}

In this appendix we discuss the numerical method
to determine the scattering matrix $S$ and
the time delay $\tau$.
The $S$-matrix of a symmetric scattering system can be
expressed in terms of the Green function $G$
\begin{eqnarray}
  \label{eq:3}
  S &=& \left(
\begin{array}{cc}
      r & t' \\
      t & r'
    \end{array}
\right),\\
  t' &=& t^T, \\
  r' &=& r,\\
  t_{\alpha\beta} &=& -i \hbar \sqrt{v_\alpha v_\beta}
  G_{\alpha\beta}(0,L), \\
  r_{\alpha\beta} &=& \delta_{\alpha\beta} - i\sqrt{v_\alpha v_\beta}
  G_{\alpha\beta}(0,0),
\end{eqnarray}
where
\begin{equation}
  \label{eq:5}
  v_\alpha = \left( \frac{2}{m}\left( E -
      \frac{\hbar^2}{2mW^2}\alpha^2 \pi^2\right)\right)^{1/2},
\end{equation}
is the velocity of mode $\alpha$ and
\begin{equation}
  \label{eq:6}
  G_{\alpha\beta}(x,x') = \int_x dy \int_{x'} dy' \phi^*_\alpha(y;x)
  \phi_\beta(y';x') G^+({\bf r},{\bf r}';E)
\end{equation}
is the projection of the retarded Green function $G^+({\bf r},{\bf
  r}';E)$ onto the local transverse modes 
\begin{equation}
  \label{eq:4}
  \phi_\alpha(y;x) = \sqrt{\frac{2}{W(x)}} \sin\left(\frac{\alpha \pi
      y}{W(x)}\right).
\end{equation}
The Green function can be calculated recursively. Expanding the
Hamiltonian
\begin{equation}
  \label{eq:7}
  H = \frac{\hbar^2}{2m} \left( \frac{\partial^2 }{\partial x^2} +
    \frac{\partial^2 }{\partial y^2} \right)
\end{equation}
in terms of the local transverse modes (\ref{eq:4}) and discretizing
in $x$ direction with a lattice constant $a$, $x=ma$, we obtain the
Hamiltonian in hybrid representation \cite{MaaZozHau94}
\begin{eqnarray}
  H_h &=&  \sum_{\alpha,m} |\alpha,m\rangle \left( \epsilon^\alpha_m + 2E_t
    \right) \langle \alpha,m| \nonumber \\
  && - \sum_{\alpha,\beta,m} \left(
    t^{\alpha\beta}_{m,m+1} |\alpha,m\rangle \langle \beta,m+1|
  \right. \nonumber \\
    && \left. +
    t^{\alpha\beta}_{m+1,m} |\alpha,m+1\rangle \langle \beta,m| \right),
  \label{eq:8}
\end{eqnarray}
with
\begin{eqnarray}
  \epsilon^\alpha_m &=& \left(\alpha \frac{a}{W(ma)}\right)^2,\nonumber\\
  t^{\alpha\beta}_{m,m+1} &=& E_t \int
  \phi^*_\alpha(y;ma)\phi_\beta(y;(m+1)a) dy,
  \label{eq:9}
\end{eqnarray}
and $E_t = \hbar^2/(2m a^2)$. In order to recursively calculate the
Green function associated with $H_h$, we split the Hamiltonian
$H_h^{M+1}$ of a system with $m=1,\ldots,M+1$ into two parts,
\begin{eqnarray}
  \label{eq:11}
  H_h^{M+1} &=& H_0 + U, \\
  H_0 &=& H_h^M + \nonumber\\
  &&\sum_\alpha |\alpha,M+1\rangle \left(
    \epsilon^\alpha_{M+1} + 2E_t \right) \langle \alpha,M+1|,
  \label{eq:15}\\
  U &=& - \sum_{\alpha,\beta} \left(
    t^{\alpha\beta}_{M,M+1} |\alpha,M\rangle \langle \beta,M+1|
  \right. \nonumber \\
    && \left. +
    t^{\alpha\beta}_{M+1,M} |\alpha,M+1\rangle \langle \beta,M|
  \right). \label{eq:16}
\end{eqnarray}
Dyson's equation,
\begin{equation}
  \label{eq:13}
  G^{M+1} = G_0 + G_0 U G^{M+1},
\end{equation}
then allows to calculate the Green function $G^{M+1}$ 
of $H_h^{M+1}$ from $G^M$ and
\begin{eqnarray}
  G_0 &=& \left( E-H_0\right)^{-1} \nonumber\\
  &=& G^M + \sum_\alpha
  |\alpha,M+1\rangle g_0^{M+1}
  \langle \alpha,M+1|,
  \label{eq:14}\\
  g_0^{M+1} &=& \left( E - \epsilon^\alpha_{M+1} -2E_t
  \right)^{-1}. \nonumber 
\end{eqnarray}
We start the recursion with $M=1$ at the left edge of the closed
billiard and iterate to the right edge at $M=N_L=L/a$. In order to
attach the leads, we again split the Hamiltonian according to
eq.~(\ref{eq:11}), but this time $H_0$ contains the Hamiltonian of the
closed billiard and the semi-infinite leads. In the leads,
$\epsilon^\alpha_m = (\alpha a/W)^2$ and $t^{\alpha\beta}_{m,m+1} =
\delta_{\alpha\beta} E_t$. $U$ is the coupling between the billiard and
the semi-infinite leads. The Green function for the leads is known
analytically \cite{Economou}.

The recursion scheme is exact for an infinite number $N_L$ of slices and
an infinite number $N_C$ of transverse modes. For numerical calculations,
both numbers have to be kept finite. We find that the
deviations from the asymptotic values for, e.g., the width $\Gamma$ of
a resonance scale as
\begin{equation}
  \label{eq:17}
  \Gamma(N_C,N_L) = \Gamma + b N_C^{-4} + c N_L^{-2},
\end{equation}
with positive numerical coefficients $b$ and $c$. For our choice of
parameters, $N=45$ transmitting modes in the leads, the values
$N_C=108$ and $N_L=12000$ give an accuracy of about $1\%$ for the
resonance width. The corrections to the position of the resonance have
the same functional form as in eq.~(\ref{eq:17}), however, they can be
either positive or negative, depending on the values of $N_C$ and
$N_L$. 
This explains the slight offset of the resonance energies with respect to the
eigenenergies of the closed system seen in Fig.~\ref{fig:paramvar}a).




\end{document}